\documentclass[twocolumn,10pt]{article}
\usepackage{epsfig}
\usefont{T1}{ptm}{m}{n}
\begin{document}

%%%%% DEFINITIONS %%%%%%%%%%%%%%%%%%%%%%%%%%%%%%%%%%%%%%%%%%%%%%%%%%%%%

\def\Xv{{\bf X}}
\def\Yv{{\bf Y}}
\def\rv{{\bf r}}
\def\vv{{\bf v}}
\def\wv{{\bf w}}
\def\av{{\bf a}}
\def\pv{{\bf p}}
\def\qv{{\bf q}}
\def\kv{{\bf k}}
\def\jv{{\bf j}}
\def\Ev{{\bf E}}
\def\Bv{{\bf B}}
\def\Av{{\bf A}}
\def\Rv{{\bf R}}
\def\Pv{{\bf P}}
\def\nv{\hat{\bf n}}
\def\grad{{\bf \nabla}}
\def\div{{\bf \nabla \cdot}}
\def\rot{{\bf \nabla} \times}
\def\xu{\hat{\bf x}}
\def\yu{\hat{\bf y}}
\def\zu{\hat{\bf z}}
\def\ie{{\it i.e.}}
\def\fv{{\bf f}}
\def\ru{\hat{\bf r}}
\def\ter{\buildrel \cdots \over}
\def\alph{\alpha_{cl}}
\def\alpharad{{2\over3}\alpha_{cl}}
\def\aT{\bf a_{\perp\hat{\bf r}}}
\def\em{electromagnetic}
\def\ni{\noindent}
\def\sumtraj{\int_{\rm traj.}}
\def\va{{\bf \dot r}_\perp}
\def\ra{{\bf      r}_\perp}
\def\w{{\rm w}} % energie du WW photon 
\def\bv{{\bf b}}
\def\qL{q_\parallel}
\def\T{_{\rm T}}
\def\vL{{\rm v}_{\rm L}}
\def\L{_{\rm L}}

%%%%%%%%%%%%% symboles speciaux %%%%%%%%%%%%%%%%%%%%%%%%%%%%%%%%%%%%

%\def\lambdabar{\lambda\raise0.4ex\hbox{\kern-0.5em\hbox{--}}\ }
\def\lambdaC{\lambda\raise0.5ex\hbox{\kern-0.5em\hbox{--}}_C\ }
\def\lesssim{{\lower0.5ex\hbox{$\stackrel{<}{\sim}$}}}
\def\gtrsim{{\lower0.5ex\hbox{$\stackrel{>}{\sim}$}}}

%%%%%%% FIN DES DEFINITIONS %%%%%%%%%%%%%%%%%%%%%%%%%%%%%%%%%%%%%%%%%%%%

%\magnification\magstep 1

\title{
SIDE-SLIPPING OF A RADIATING PARTICLE
} 

\author{
{\bf 
X. Artru$^a$, G. Bignon$^b$, T. Qasmi$^c$
} \\
{\small $^a$
Institut de Physique Nucl\'eaire de Lyon, 
IN2P3-CNRS \& Universit\'e Claude-Bernard, France} \\
{\small $^b$
Ecole Normale Sup\'erieure de Lyon, France} \\
{\small $^c$ deceased} \\
}
\date{
{\begin{flushleft} \normalsize
%(abstract)
Radiation reaction is revisited, first in a new classical aproach, 
where the physical particle 4-momentum is redefined as the 
energy-momentum flux across the future light cone and is not 
parallel to the 4-velocity.  
Then in a semi-classical approach, it is shown that,
when emitting a photon, the particle "side-slips" 
transversaly to its initial momentum, justifying the non-colinearity
between momentum and mean velocity.
Side-slipping is finally checked in a pure quantum mechanical 
treatment of synchrotron radiation. \\
~~PACS: 3.50.De ~ 41.60.-m ~ 61.85
\end{flushleft}
}}

\maketitle

%--------------------------------------------------------------------------

\ni
{\bf 1. RECALL ABOUT RADIATION} 

An electron submitted to an external field $(\Ev_{in}, \Bv_{in})$
in vacuum emits radiation with the power~:
$$
{d \,W_{rad} \over dt} = {2\over3} \alph \ \left({d\Xv \over dt}\right)^2
\eqno(1)
$$
The relativistic 4-vector generalization is 
$$
{d \, P^\mu_{rad} \over d\tau} = 
{2\over3} \alph \ (\ddot X \cdot \ddot X) \ \dot X^\mu 
\eqno(1')
$$
We take unified dimensions for space and time ($c=1$) with the
$( - + + + )$ metric. $\dot \Xv \equiv d\Xv/d\tau$, 
where $\tau$ is the proper time.
We also use rationalized Maxwell equations, e.g. $\div \Ev = \rho$.
We keep $\hbar \ne 1$ and define $\alph \equiv e^2 / (4\pi) = \hbar/137$,
where $e = -|e|$ is the charge of the electron.  
To account for the loss of the electron energy,
Abraham and Lorentz introduced the dissipative force
$$
\fv_{reac} = {2\over3} \alph \ \ter \Xv
\eqno(2)
$$
The non-relativistic equation of motion is then
$$
m \ddot \Xv = e \ (\Ev_{in} + \dot \Xv \times \Bv_{in}) + \fv_{reac}
$$
whose relativistic generalization is the Abraham-Lorentz-Dirac (ALD)
equation~:
$$
m \, \ddot X  =  e \ F_{in}(X) \cdot \dot X 
- {2\over3} \alph \, [\,  (\ddot X \cdot \ddot X)\, \dot X - {\ter X} \,]
\eqno(3)
$$
We use the notation 
$(F \, \dot X)^\mu \equiv F^{\mu\nu} \ \dot X_\nu$.
$F_{in} = \{ \Ev_{in}, \Bv_{in} \}$ is the "incoming" 
(or "external") electromagnetic field, 
related to the total, retarded, advanced and outgoing fields by
$$
F_{tot} = F_{in} + F_{ret} = F_{adv} + F_{out}
\,,
\eqno(4)
$$\
$$
F_{rad} = F_{out} - F_{in} = F_{ret} - F_{adv}
\,. 
\eqno(4')
$$
In the following we shall omit the suffix $in$.
An excellent review on radiation reaction 
can be found in Ref.[1]

\medskip
\ni
{\it The "mad electron".}

Although mathematically elegant, the ALD equation is not physically 
acceptable for the following reasons~:

* a third initial condition $\ddot X(0)$ is needed in addition
to $X(0)$ and $\dot X(0)$.

\par * for almost every $\ddot X(0)$, the electron eventually goes
into a {\it run-away} motion.
%, i.e., it is self-accelerated indefinitely.

\par * given $X(0)$ and $\dot X(0)$, there may exist one
(or a discrete set of) $\ddot X(0)$ such that the electron 
avoids run-away motion, but this value depends 
on all the fields $F_{in}(X)$
that the electron will encounter in the future. 
Saying that "nature precisely chooses this $\ddot X(0)$" constitutes 
a violation of the causality principle.

\ni
One may compare this situation with the following one~:
In a bus, a passenger puts a stick vertical on the floor
and wants it to remain standing up in equilibrium during the whole journey,
and also after the bus has stopped.
To counter-act the accelerations of the bus, 
he or she must give some initial angular velocity to the stick (Fig.1).
To do so, the passenger must know exactly in advance
the accelerations of the vehicle during the whole journey.

The run-away instability is probably related to the point-like limit
of the classical electron considered by Lorentz~:
For a sufficiently small radius, the electrostatic self-energy 
is larger than the physical mass. Then the electron "core" 
has a negative mass and "likes" to accelerate, 
since that lowers its kinetic energy.

It is possible to find approximations of the ALD equation, 
valid to first order in $\alph$, which remove the arbitrariness
of $\ddot X(0)$ and have no run-away solutions.  
One of them [2] is obtained by replacing
$\ddot X$ and ${\ter X}$ in the right-hand side of (3) by their
values calculated without radiation reaction,
$$
\ddot X \quad \longrightarrow \quad (e/m) \ F \dot X 
\,,
$$
$$
{\ter X}  \quad \longrightarrow \quad
 (e/m) \ \dot F \dot X 
 + (e/m)^2 \ F F \dot X  
$$
with $\dot F = \dot X^\lambda \ \partial_\lambda F(X)$.
On obtains
$$
m \ddot X = e F \dot X
+ \sigma_{Th}
\left[ 
F  F  \dot X - 
(F \dot X)^2  \dot X
\right]
+ {2e\over3} r_{cl}
 \dot F \dot X
\,,\eqno(5)
$$
where 
$r_{cl} = e^2 / (4\pi m)$ is the classical electron radius 
and 
$\sigma_{Th} = (8\pi / 3) r_{cl}^2$ 
the Thomson cross section. 
The second term of (5) can be interpreted 
as the radiation pressure of the incoming field.
%In this paper also, we shall not try to solve the run-away problem
%and work only to first order in $\alph$. 

%In this paper, we will derive the ALD equation in a  

\bigskip
\ni
{\bf REFORMULATION OF THE ALD EQUATION AND NEW APPROXIMATIONS}

Usually, one identifies $m \ddot X$ with the physical 4-momentum 
of the particle.
Then Eqs. (1') and (3) do not seem to conserve the total 4-momentum 
instantaneously, because of the {\it Schott term} 
$ (2\alph/3) \,  {\ter X} $.
Redefining the 4-momentum as
$$
P^\mu = m \, \dot X^\mu - \alpharad \, \ddot X^\mu
\,,
\eqno(6)
$$
the ALD equation can be replaced by the following system~:
%
%\begin{equation}
%\begin{cases}
$$
m \, \dot X = P + \alpharad \, \ddot X
\eqno(7a)
%& \text{(7a)} \\
$$
$$
\dot P = e \ F \ \dot X 
- \alpharad \ (\ddot X \cdot \ddot X) \ \dot X 
%& \text{(7b).}
\,.\eqno(7b)
$$
%\end{cases}
%\end{equation}
%
Eqs.(1') and (7b) make the instantaneous conservation 
of the total 4-momentum manifest. 
On the other hand, the mass is not conserved~: 
$P \cdot P \ne -m^2$ and $\Pv$ is not colinear 
to the electron velocity. 
These two features are not physically damning. 
(6) can be approximated by
$P(\tau) \simeq m \, \dot X(\tau - 2r_e/3)$,
telling that the electromagnetic part of $P^\mu$ 
follows the variations of the core velocity with some delay.

%%%%%%%%%%%%%%%%%%%%%%%%%%%%%%%%%%%%%%%%%%%%%%%%%%%%%%%%%%%%%%%%%%%%%%%%
\begin{figure}
\centering\epsfig{file=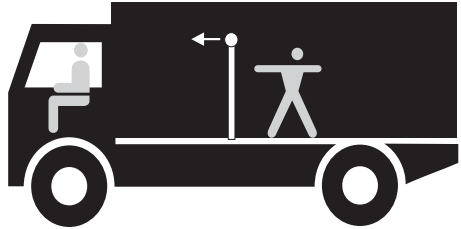,width=6cm,height=3cm}
\caption{Fig.1. Stick standing in equilibrium in a truck.}
%\label{fig:boxhb}
\end{figure}
%%%%%%%%%%%%%%%%%%%%%%%%%%%%%%%%%%%%%%%%%%%%%%%%%%%%%%%%%%%%%%%%%%%%%%%%%

In what follows, we shall show that (6) is a quite natural 
definition of $P^\mu$. 
As usual, we separate $P^\mu$ in core and electromagnetic contributions~:
$$
P = m_c \dot X + \delta P
\eqno(8)
$$
with
$$
\delta P^\mu = \int_\Sigma d\Sigma_\nu \ \Theta^{\mu\nu}
\,,
\eqno(9)
$$
where $\Theta^{\mu\nu}$ is the energy-momentum flux tensor
of the electron field. 
The latter field is not uniquely defined~: 
it can be the retarded one, the advanced one or any linear combination
of the two. 
Looking at the first decomposition of Eq.(4),
we choose the {\it retarded} field. 
So we consider that the incoming field does not contribute
to the self 4-momentum and only exerts a force on the core 
according to the first term of (3). 
As hyper-surface $\Sigma$ we a priori choose the future light cone 
of the electron (Fig.2). This avoids a contribution from
the {\it radiated} field $F_{rad}$, the 4-momentum of which flows parallel
to the cone and does not cross it. 

%%%%%%%%%%%%%%%%%%%%%%%%%%%%%%%%%%%%%%%%%%%%%%%%%%%%%%%%%%%%%%%%%%%%%%%%
\begin{figure}
\centering\epsfig{file=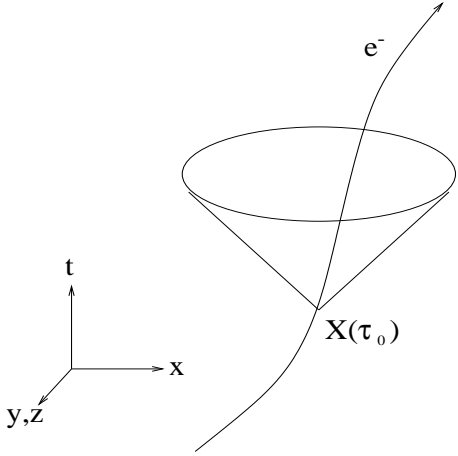,width=6cm,height=6cm}
\caption{World line and future light cone of the particle.}
%\label{fig:boxhb}
\end{figure}
%%%%%%%%%%%%%%%%%%%%%%%%%%%%%%%%%%%%%%%%%%%%%%%%%%%%%%%%%%%%%%%%%%%%%%%%%

Let us consider a point $X^\mu(\tau_0)$ of the electron world line.
For a space-time point $Y^\mu$ of its future light cone, we define
$$
Y - X(\tau_0) \equiv R = (r, \rv)
\ = r \, (1, \ru)
\,.
$$
The integrand of (9) can be evaluated most easily in the electron
rest frame, using standard formulas for $\Ev_{ret}$, $\Bv_{ret}$,
$\Theta^{\mu\nu}$ and 
$$
(d\Sigma_\nu) = d^3\rv \, (1, - \ru) 
%= - {d^3\rv \over r} \ R_\mu
\,.
\eqno(10)
$$
We only give the result~:
$$
\delta P = {\alph\over8\pi} \int {d^3\rv \over r^4} \ (1, \ru)
$$
$$
=  {\alph\over8\pi} \int 
{d^3\rv \over r} \ [\dot X(\tau_0) \cdot R]^{-4} \ R
\,.
\eqno(11)
$$
The second expression is Lorentz invariant
and applies as well in frames where the electron is not at rest.
Note that the acceleration does not enter this formula. 
It confirms that there is no contribution of the radiated field.

The integral diverges at $\rv=0$, recalling that the classical 
self-energy of a point-like charge is infinite.
In the following, we will assume that the electron has some
very small but finite extension $r_c$.
To treat the divergence, we truncate the light cone by a hyperplane 
orthogonal to the electron world line at $X(\tau)$
where $\tau - \tau_0 = \rho$ is a small distance, but larger than $r_c$ (Fig.3). 
We now take the rest frame of the electron at $X(\tau)$ (not $\tau_0$).
We close the cone, truncated at $r \simeq \rho$,
by the piece of hyperplane
$R^0 = \rho$, $|\Rv| \le \rho$
and integrate (9) on the new hypersurface 
(in grey on Fig.3) which we call a "flower-pot".
To first order in $\rho$, 
$$
\dot X(\tau_0) = (1, - \rho \ \ddot \Xv) 
$$
$$
[\dot X(\tau_0) \cdot R]^{-4} 
= r^{-4} \ [\, 1 - 4 \rho \ \ddot \Xv \cdot \ru \, ]
\,.
\eqno(12)
$$
%

%%%%%%%%%%%%%%%%%%%%%%%%%%%%%%%%%%%%%%%%%%%%%%%%%%%%%%%%%%%%%%%%%%%%%%%%
\begin{figure}
\centering\epsfig{file=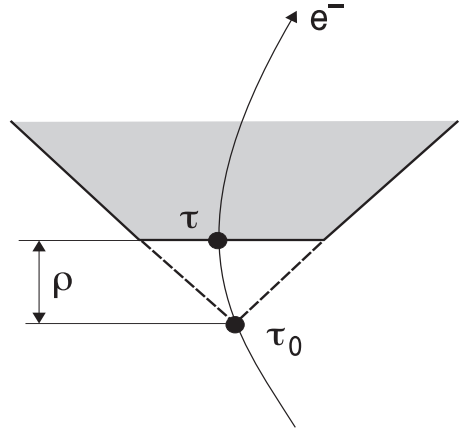,width=6cm,height=6cm}
\caption{Truncated lightcone or "flower-pot".}
%\label{fig:boxhb}
\end{figure}
%%%%%%%%%%%%%%%%%%%%%%%%%%%%%%%%%%%%%%%%%%%%%%%%%%%%%%%%%%%%%%%%%%%%%%%%%

The truncated integral of (11) is
$$
\delta P_{r>\rho} 
= \alph \left( {1\over2 \rho} , - {2\over3} \ddot \Xv(\tau) \right)
$$
$$
= {\alph\over 2 \rho} \, \dot X(\tau) 
- {2\over3}\alph \, \ddot X(\tau)
\,,
\eqno(13)
$$
the second expression being frame-independent.
The hyperplane piece (bottom of the flower-pot) 
is the interior the sphere of radius $\rho$
at fixed time. It is approximately centered at $\Xv(\tau)$,
the displacement being of second order in $\rho$.
Its contribution to (9) is
$$
\delta P_{r<\rho} = (\delta m_{r<\rho}, 0)
= \delta m_{r<\rho} \, \dot X(\tau)
\,.
\eqno(14)
$$
The total 4-momentum at proper time $\tau$ (not $\tau_0$) is
obtained from (8), (13) and (14). We recover the new definition
$$
P(\tau) =  m \, \dot X(\tau) - {2\over3}\alph \, \ddot X(\tau)
\eqno(6)
$$
where
$$
m = m_c + \delta m_{r<\rho} + {\alph\over 2 \rho}
\eqno(15)
$$
is the renormalized mass of the electron. The third term is
the Coulomb self-energy at $r \ge \rho$ whereas the detailed
short range structure of the electron is summarized in the sum
of the first two terms. 

Let us make the energy-momentum balance in the space-time region between
two successive "flower-pots" at proper times $\tau$ and $\tau + d \tau$~:
\par *
$P(\tau)$ is coming through the first flower-pot,
\par * 
$dP_{in} = e \ F_{in} \, \dot X \ d\tau$ 
is brought to the core by $F_{in}$,
\par * 
$dP_{rad} = {2\over3}\alph \ (\ddot X \cdot \ddot X) \ \dot X \ d\tau$
is radiated at infinity between the two flower-pots,
\par *
$P(\tau+d\tau) = P(\tau) + \dot P \ d\tau$
is outgoing through the second flower-pot.

\ni 
Adding the first two quantities and subtracting the last two ones
must give zero. This gives (7b).

The above calculations constitute a new and relatively simple
derivation of the ALD equation, written in the form (7). 
From this form on can derive new types of approximations [3-5], 
also valid to first order in $\alph$. The simplest one to implement
in a computer code is obtained replacing
$\ddot X$ in the right-hand sides by $(e/m^2) \, F P$.
On may in addition replace the last $\dot X$ by $P/m$. 
Compared to (5), these approximations have the advantage 
of not involving the field derivatives.

\bigskip

\ni
{\bf 3. SEMI-CLASSICAL APPROACH}

Eq.(6) tells that the momentum does not follow the velocity,
but one may see things the other way around and say that the electron
does not follows the direction of its momentum.
We call this phenomenon side-slipping, by analogy with 
a skier whose track is not always tangential to the skis (Fig.4)

%%%%%%%%%%%%%%%%%%%%%%%%%%%%%%%%%%%%%%%%%%%%%%%%%%%%%%%%%%%%%%%%%%%%%%%%
\begin{figure}
\centering\epsfig{file=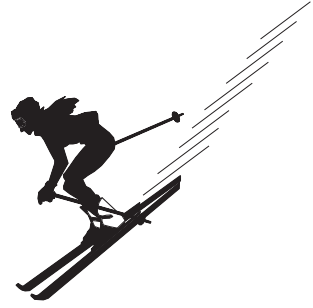,width=4cm,height=4cm}
\caption{Side-slipping skier.}
%\label{fig:boxhb}
\end{figure}
%%%%%%%%%%%%%%%%%%%%%%%%%%%%%%%%%%%%%%%%%%%%%%%%%%%%%%%%%%%%%%%%%%%%%%%%%

A discrete side-slipping is naturally obtained 
in a semi-classical description of the process
$e^- \to e'^- + photon$ in an external field.
If we consider this process as instantaneous
and local at a definite point $\Xv$ of the trajectory, 
it cannot satisfy the conservation of both momentum,
\def\en{\epsilon}
$$
\Pv = \Pv' + \hbar \, \kv
\eqno(16)
$$
and energy
$$
\en = \en' + \hbar\omega
\eqno(17)
$$
with $\en\equiv\sqrt{\Pv^2+m^2}$.
However, 4-momentum conservation becomes possible
if we assume that the final electron trajectory 
starts from a point $\Xv'\ne\Xv$ aside from the initial trajectory. 
In the case of a static electric field, we replace (17) by
$$
U(\Xv) + \en = U(\Xv') + \en' + \hbar\omega 
\eqno(17')
$$
where $U(\Xv)$ is the potential energy. (17') and (17) give
$$
\delta U = U(\Xv') - U(\Xv) \simeq
% - \hbar \ (\omega - \vv \cdot \kv)
- {\en \over \en'} \hbar\omega 
\, {\gamma^{-2} + \theta^2 \over 2}
\,,
\eqno(18)
$$
where $\theta$ is the angle between $\Pv$ and $\kv$. 
$\delta U$ is obtained by a finite displacement 
$\delta\Xv = \Xv' - \Xv$ of the electron toward a lower potential energy.
In the ultrarelativistic case, we take $\delta\Xv$ perpendicular
to the trajectory~:
$$
\delta\Xv = 
- {\en \over \en'} \hbar\omega 
\, {\gamma^{-2} + \theta^2 \over 2}
\, {\fv_\perp\over|\fv_\perp|^2}
\eqno(19)
$$
where $\fv_\perp$ is the transverse component of the force.
Such a "side-slipping" was already introduced in channeling radiation 
(Eqs.15-17 of Ref.[6]). 
It contributes to the decrease of the transverse energy
which explains the very fast energy loss of axially channeled electrons 
above hundred GeV. 

\medskip

Let us now consider synchrotron radiation
in a uniform magnetic field $\Bv = - B \, \zu$
derived from the vector potential
$$
\Av = (y B, 0 , 0)
\,.
\eqno(20)
$$
The particle hamiltonian is $(\Pv^2 + m^2)^{1/2}$ where
$\Pv = \pv - e \Av$ is the mechanical momentum and $\pv$ the canonical one.
In the gauge (20), the hamiltonian is invariant under translation 
in the $x$ and $z$ directions, therefore $p_x$ and $p_z$ are conserved.
We assume that the photon is emitted when the electron is at  
$x=0$, $y=R$ (Fig.5). Then we require the conservation laws
(16-17), but with $\pv$ and $\pv'$ in place of $\Pv$ and $\Pv'$.
For the $x$-component it writes
$$
P_x + e y B  = P'_x + e y' B  + \hbar \, k_x
\,,
\eqno(21)
$$
where have anticipated a side-slipping $y' = y + \delta y$.
For $|e|B\delta y$ we obtain the same result (18) as for $\delta U$
and, since $|e|B \simeq |\fv|$, $\delta y$ is given by (19) or 
$$
\delta y = - {\hbar\omega \over \en'} \, R
\, {\gamma^{-2} + \theta^2 \over 2} 
\,.\eqno(22)
$$
%

%%%%%%%%%%%%%%%%%%%%%%%%%%%%%%%%%%%%%%%%%%%%%%%%%%%%%%%%%%%%%%%%%%%%%%%%
\begin{figure}
\centering\epsfig{file=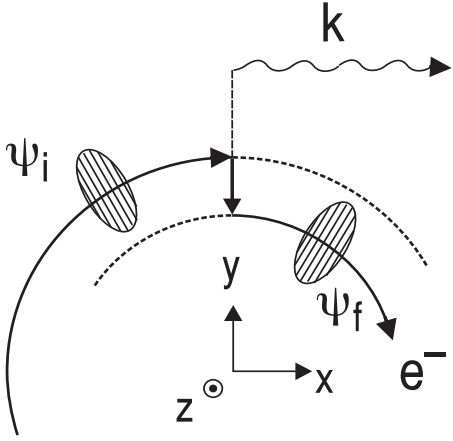,width=6cm,height=6cm}
\caption{Photon emission in a synchrotron.}
%\label{fig:boxhb}
\end{figure}
%%%%%%%%%%%%%%%%%%%%%%%%%%%%%%%%%%%%%%%%%%%%%%%%%%%%%%%%%%%%%%%%%%%%%%%%%

The side-slipping has also the virtue of insuring 
angular momentum conservation. Let us consider again the 
circular trajectory of Fig.5, but now due to the spherically 
symmetric potential $U(|\Xv|)$. 
Neglecting spin, the $z$-component of
the angular momenta of the initial and final electrons are
$$
L_z = - y \ P_x
\,,\quad 
L'_z = - y' \ P'_x
\,.
\eqno(23)
$$
Here we neglect the quantum recoil effect, 
i.e. we use the classical or {\it soft photon} approximation 
($\hbar\omega \ll \en - m$).
The source of the radiation - and the radiation itself - is invariant 
under a time translation by $\Delta t$ times a rotation
by the angle $v\Delta t/R$. 
For a photon quantum state of definite angular momentum $J_z$ 
and frequency $\omega$, this invariance is expressed as 
$$
\exp[- i (v\Delta t /R) J_z] \times \exp(- i \, \omega \, \Delta t) = 1
$$
therefore
$$
J_z = - \omega \, R /v
\,.
\eqno(24)
$$
The conservation of angular momentum, 
$$
L_z = L'_z + J_z
\,.
\eqno(25)
$$
together with that of linear momentum along $\xu$,
$$
P_x = P'_x + \hbar k_x
\eqno(26)
$$
yield the result (22) again, with $\en\simeq \en'$.

Incidentally, identifying (24) with the "classical photon" result
$J_z = - y k_x$ implies a "side-slipping" for the photon also~:
$$
y_{phot} - R = {\gamma^{-2} + \theta_z^2 \over 2} \, R
\,,
\eqno(27)
$$
which could be observed at low-energy synchrotron machines.

%%%%%%%%%%%%%%%%%%%%%%%%%%%%%%%%%%%%%%%%%%%%%%%%%%%%%%%%%%%%%%%%%%%%%

The side-slipping formula (19) can be generalized in a covariant form,
writing the 4-momentum conservation as
$$
P + Q = P' + \hbar K
\eqno(28)
$$
We assume that $Q$ is provided by the work of the 
external field along $\delta X$~:
$$
Q = e \ F \, \delta X
\eqno(29)
$$
Squaring the two sides of (28), using $P^2 = P'^2 = - m^2$, $K^2=0$
and neglecting $Q^2$ a priori, we obtain
$$
P \cdot Q = \hbar K \cdot P' 
\quad \left( \simeq {\en \over \en'} \, \hbar K \cdot P \right)
\,.
\eqno(30)
$$
Using $P \cdot F \dot P = - \dot P \cdot F P$ and,
to first order in $\alph$, $P = m \dot X$,  
$F \dot X = \dot P$, 
one can verify that
$$
\delta X^\mu = {\hbar\over m} \, {- K \cdot P' \over \dot P \cdot \dot P} 
\ \dot P^\mu
\,,
\eqno(31)
$$
inserted in (29), satisfies (30).
%In the soft photons approximation, $\hbar \omega \ll \en$,
%we can replace $P'\cdot K$ by $P \cdot K$.
The neglect of $Q^2$ has to be checked a posteriori from (29).
We expect it to be small if the external field varies smooththly,
e.g. in synchrotron or channeling radiations,
interpreting $Q$ as the momentum of the virtual photon(s)
taken from the external field.

In the limit $\hbar \to 0$, the 4-momenta of the individual photons
goes to zero and their number goes to infinity
so that the total radiated 4-momentum is finite and given by (1').
Summing all the small side-slippings (31) during the proper 
time $d\tau$, approximating $K\cdot P'$ by $K \cdot P$,
one recovers Eq.(7a),
to first order in $\alph$. This is illustrated in Fig.6.

%%%%%%%%%%%%%%%%%%%%%%%%%%%%%%%%%%%%%%%%%%%%%%%%%%%%%%%%%%%%%%%%%%%%%%%%
\begin{figure}
\centering\epsfig{file=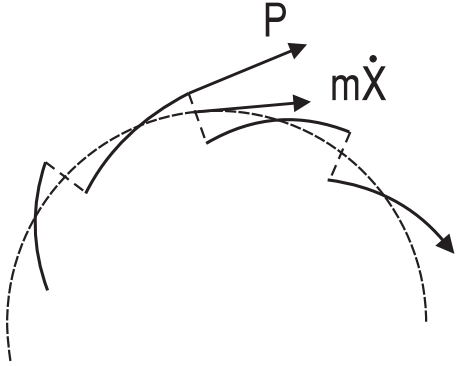,width=6cm,height=5cm}
\caption{Semi-classical electron trajectory emitting photons
successively.}
%\label{fig:boxhb}
\end{figure}
%%%%%%%%%%%%%%%%%%%%%%%%%%%%%%%%%%%%%%%%%%%%%%%%%%%%%%%%%%%%%%%%%%%%%%%%%

\bigskip %%%%%%%%%%%%%%%%%%%%%%%%%%%%%%%%%%%%%%%%%%%%%%%%%%%%%%%%%%%%55

\ni
{\bf 4. FULL QUANTUM DERIVATION}

\def\dx{\partial_x}
\def\dy{\partial_y}
\def\dz{\partial_z}
\def\dt{\partial_t}
\def\pl{p}

Side-slipping was deduced above from semi-classical arguments of
energy, momentum and angular momentum conservation.
Here we will derive side-slipping from a full quantum treatment, 
in the particular case of synchrotron radiation. 
Neglecting electron spin, we start from the Klein-Gordon equation 
(now $\hbar=1$),
$$
[(\grad - ie \Av)^2 - \dt^2 - m^2] \, \Psi = 0 
\eqno(32)
$$
and consider a wave packet of the form
$$
\Psi = e^{i \pl x -i \en t} \, \psi
\eqno(33)
$$
where $P^\mu = (\en,\pl,0,0)$ is a reference 4-momentum
and $\psi$ a slowly varying function of $(t,x,y,z)$.
Using $\en^2 = \pl^2 + m^2$, (32) becomes
$$
[i \en \partial_+ - \partial_+ \partial_- + i(E-\pl) \dx
+ \dy^2 + \dz^2
$$
$$
+ 2fy (\pl - i\dx) - f^2 y^2]
\, \psi = 0
\eqno(34)
$$
where $\partial_{\pm} = \dt \pm \dx$ and $f=|e|B$.
Assuming $\en \simeq \pl \gg m$ we consider
$\partial_+$ to be of order $\en^{-1}$, 
which allows us to neglect the second and third terms of the square bracket.
Furthermore, we take a wave packet located near $(x,y,z)=0$ at time $t=0$ 
(we change the origin of the coordinates in Fig.5). 
So we neglect the terms in $y^2$ and $y\dx$ (but not in $yp$).
We get
$$
[i(\dt+\dx) + {1\over2\en} (\dy^2 + \dz^2) -  fy]
\, \psi = 0
\,.
\eqno(35)
$$
Looking for solution of the form
$$
\psi(t,x,y,z) = \chi(x-t) \phi(t,y,z)
\,,
\eqno(36)
$$
we are left with the 2-dimensional Schr\"odinger equation 
for a particle of mass $\en=\gamma m$ in the linear potential 
$V(y) = by$~:
$$
[i \dt + {1\over2\en} (\dy^2 + \dz^2) - fy]
\, \phi = 0
\eqno(37)
$$
Using the coordinate of the accelerated frame
\def\ya{y_a}
$$
\ya = y + {f\over2\en} t^2
\eqno(38)
$$
%
%with $a = f/\en$ 
and setting
$$
\phi(t,y,z) = \phi_a(t,\ya,z) 
\ \exp\left( -ifty -i{b^2 t^3 \over 6\en} \right)
\eqno(39)
$$
we transform (37) in the free-particle Schr\"odinger equation,
$$
[i \dt + {1\over2\en} (\partial_{\ya}^2 + \dz^2)]
\, \phi_a(t,\ya,z) = 0
\,.
\eqno(40)
$$
Thus $\phi_a$ can be expanded in plane waves~:
$$
\phi_a = \int\!\!\int {dq\over2\pi}{dr\over2\pi}
\tilde\phi_a(q,r) \
\exp\left(
iq\ya+irz-i{q^2+r^2\over2\en}t
\right) 
\eqno(41)
$$
To sum up,
$$
\Psi = e^{i \pl x -i \en t}
\chi(x-t) 
\exp\left(
-ifty-i{b^2 t^3 \over 6E}
\right)
\int\!\!\int {dq\over2\pi}{dr\over2\pi}
$$
$$
\tilde\phi_a(q,r)
\exp\left[i\left(
qy+{qft^2\over2\en}+rz-{q^2+r^2\over2\en}t 
\right) \right]
\eqno(42)
\,.
$$

\def\pol{{\bf a}}

We consider the transition from the electron state 
$
|i\rangle = |\Psi\rangle
$
(in the Schr\"odinger representation)  
to the electron + photon state 
$
|f\rangle = |\Psi'\rangle \otimes |\kv,\pol\rangle
$
where $\Psi'$ is given by (42) with primed quantities.
The wave packets $\Psi$ and $\Psi'$ are represented by
the striated ellipses of Fig.5.
Taking (without loss of generality) $\kv$ along the $x$ axis,
the photon vector potential is given by
$$
\Av(t,\Xv) = \pol \, e^{i k (x - t)}
\,.
\eqno(43)
$$
To first order in perturbation, the transition amplitude is 
$$
\langle f|S|i\rangle = 
-i \int dt \langle f|H_I|i\rangle 
\eqno(44)
$$
with the interaction hamiltonian $H_I$ given by 
$$
\langle f|H_I|i\rangle = ie \int d^3\Xv   
\ \Psi'^*(t,\Xv) 
\ \Av^*(t,\Xv)\cdot\grad 
\Psi(t,\Xv)
\,.
\eqno(45)
$$
We now combine Eqs.(42-45). 
Integrations over $y$ and $z$ impose $q = q'$ and $r = r'$.
Using the shifted variable $x'=x-t$, the integration over the 
$x$-dependent factors gives 
$$
e^{i(p'+k-p)t}
\int dx' \ e^{i(p'+k-p)t} \ \chi(x') \ \chi'^*(x')
\,.
\eqno(46)
$$
We introduce the parameter
$$
\Lambda = (\en' - \en - p' + p) / m^2
\simeq 
{k \over 2 \en \en'}
\eqno(47)
$$
and write the remaining 3-fold integral as
$$
I= \int\!\!\int\!\!\int
dt\,dq\,dr \ 
\tilde\phi'^*_a(q,r)
(q\cdot\pol^*_y + r\cdot\pol^*_z)
\tilde\phi_a(q,r)
$$
$$
\exp\left\{i \Lambda \left[
(m^2 + q^2 + r^2) t - qft^2 + f^2 {t^3  \over 3}
\right]
\right\}
\,,
\eqno(48)
$$
Shifting the time variable $t'=t-q/f$,
we can decouple the exponential into 
$$
\exp\left\{i \Lambda \left[
(m^2 + r^2) t' + f^2 {t'^3  \over 3}
\right]
\right\}
\eqno(49)
$$
and
$$
\exp\left\{i {\Lambda\over f} \left[
(m^2 + r^2) q + {q^3  \over 3}
\right]
\right\}
\,.
\eqno(50)
$$
The phase factor of (49) is the same as 
in the semi-classical radiation formula,
$$
\exp\left\{ i {\en\over\en'} 
[\omega t' - \kv\cdot\Xv(t')]
\right\}
\,,
\eqno(51)
$$
knowing that the transverse components of the velocity $d\Xv/dt'$ are
$v_y(t') = - ft'/\en$, $v_z=r/\en$.
The factor $\en/\en'$ is a recoil correction.

The factor which interests us is (50).
Linearizing the cubic term about the mean value $<q> = \en v_y$
and replacing $r$ by $<r> = \en v_z$, we can rewrite (50) as
$$
C \cdot \exp(-iq\,\delta y) 
\eqno(52)
$$
where $\delta y$ is equal to the right-hand side of (19) or (22).
In (52) we recognize the operator of the $y$-translation by $\delta y$,
written in the momentum space representation. 
The maximum transition amplitude is obtained when the wave packet
$\tilde\phi'_a$ is transversaly shifted
from $\tilde\phi_a$ by $\delta y$. 
This confirms the semi-classical derivation of the side-slipping.

\bigskip %%%%%%%%%%%%%%%%%%%%%%%%%%%%%%%%%%%%%%%%%%%%%%%%%%%%%%%%%%%%55

\ni
{\bf 5. CONCLUSION}

In this study, we have got new insight in the radiation mechanism. 
Using purely classical, semi-classical and quantum-mechanical approaches, 
we have shown that the velocity and the properly defined momentum
of the radiating particle are not parallel, as illustrated in Fig.6.
The classical run-away problem still remains unsolved, 
but we have obtained a new approximation of the ALD equation,
without run-away and not involving the field derivatives. 
It can be easily implemented on a computer code.

The discrete side-slipping accompanying the emission of a photon
is of the order of the comptom wavelength, hence hardly detectable.
However its contribution to the decrease of the transverse energy
of a high-energy electron channeled in crystals may be non-negligible.
The "side-slipping of the photon" (27), much larger than the electron one,
may be observed with precise optics.

The transverse jumping of the particle from the initial
to the final  trajectory has no classical counterpart. 
It can be viewed as a tunnel effect. 
A similar effect should take place in the crossed reaction 
$\gamma \to e^+ + e^-$ in a strong field (Eq.2 of Ref.[7]).

\medskip
\ni
Part of this work was supported by the INTAS contract 97-30392~:
"Theoretical Investigation of Propagation of Particles, Ions and 
X-Rays through Straight and Bent Nanotubes and Associated Phenomena".
Two of us (G.B. and T.Q.) took part in this work during training periods 
at Institut de Physique Nucl\'eaire de Lyon.

\medskip
\centerline{\bf REFERENCES}

{1.} K.T. McDonald (2000), Limits of the applicability of classical
electromagnetic fields as inferred from the radiation reaction,
{\it ArXiv:physics}/0003062.

{2.} L.D. Landau and E.M. Lifshitz, 
{\it Course of Theoretical Physics},
{\bf 2}, {\it Classical Theory of Fields}, 1975, Pergamon.

3. T. Qasmi, Rayonnement de canalisation dans les nanotubes
(Training work report, 1998, unpublished).

4. G. Bignon, Force de r\'eaction au rayonnement
(Training work report, 2001, unpublished).

5. X. Artru, G. Bignon, A semi-classical approach to the radiation
damping force 
(NATO Advanced Research Workshop
on Electron-Photon Interaction in Dense Media, Nor Hamberd, Armenia,
25-29 June 2001) - submitted to {\it NATO Science series 2}.

{6.} X. Artru, Self-amplification of channeling radiation
of ultrarelativistic electrons due to loss of transverse energy,  
{\it Phys. Lett.}, 1988, v. A128, p.302-306. 

{7.} X.Artru et al, Observation of channeling and blocking effect
in pair creation in a Ge crystal,  
{\it Phys. Lett.}, 1993, v. B313, p.483-490.

\end{document}